\documentclass[letterpaper]{jpconf}
\usepackage{graphicx,citesort,collref,url}
\newcommand{\met}{$E\!\!\!/_T$}
\newcommand{\dzero}  {\hbox{DO\kern-0.62em\raise+0.2ex\hbox{/}}}

\begin{document}
\nocite{*}
\title{Latest Electroweak Results from CDF}

\author{Mark Lancaster on behalf of the CDF collaboration}

\address{Department of Physics and Astronomy, UCL, Gower Street, London WC1E 6BT, UK.}

\begin{abstract}
The latest results in electroweak physics from proton anti-proton collisions at the Fermilab Tevatron recorded by the 
CDF detector are presented. The results provide constraints on parton distribution functions, the mass of the Higgs 
boson and beyond the Standard Model physics.
\end{abstract}

\section{Introduction}
Electroweak physics studies the properties and interactions of the electroweak gauge bosons: the $W$, $Z$ and 
photon. $W$ and $Z$ particles are being produced at a prodigious rate in proton anti-proton collisions at a centre of 
mass energy of 1.96~TeV at the Fermilab Tevatron. The CDF detector has already recorded around 6 million $W$ 
and 600,000 $Z$ bosons and before the end of data-taking in 2011 will have accumulated samples that will be twice 
those anticipated from the first LHC running period (1 fb$^{-1}$ of integrated luminosity at $\sqrt{s}$ = 7~TeV). These large 
data samples allow the electroweak sector to be probed to high accuracy. Measurements of the rapidity distributions of 
$W$ and $Z$ bosons are providing precise constraints on the parton distribution functions (PDFs) and a study 
of di-boson production ($WW$, $ZZ$, $WZ$, $W\gamma$, $Z\gamma$) and searches for rare decay modes e.g. $W 
\rightarrow \pi\gamma$ are providing stringent constraints on physics beyond the Standard Model (SM). A 
precise measurement of the mass, $M_W$, of the $W$ boson (in conjunction with that of the top-quark) allows one 
to predict the mass of the Higgs boson which presently points to a light Higgs boson.
The results presented in the following sections are from 3--5~fb$^{-1}$ of integrated luminosity. At the time of writing 
CDF is analysing datasets of $\sim$~7~fb$^{-1}$ and is expected to accumulate $\sim$~10~fb$^{-1}$ before the 
scheduled end of the Tevatron programme in September 2011 and as such significant further improvements in the precision and scope with 
which the electroweak sector can be probed are expected.

\section{PDF Constraints}
Precise PDFs are necessary to make robust predictions of the SM background and new physics at all colliders, 
but particularly the LHC. The 
LHC will ultimately provide some of the measurements that constrain PDFs but at present the constraints predominantly 
come from HERA (and fixed-target) deep-inelastic data, HERA/Tevatron jet-data and Drell-Yan data both from the 
Tevatron and at lower centre of mass energy experiments. The rapidity, $Y_Z$, of the $Z$ from a Drell-Yan event is 
given by $Y_Z = 0.5 \ln \left( \frac{x_p}{x_{\overline{p}}}\right)$ where $x_p(x_{\overline{p}})$ is the momentum fraction of the 
quarks in the proton (anti-proton) participating in the Drell-Yan process. At large values of $|Y_Z|$ one thus simultaneously probes 
the high-$x$ and low-$x$ structure of the proton which are the regions of greatest PDF uncertainty.  The $u$-valence 
distribution is already well constrained by $F_2$ data and consequently the latest CDF $Z$ rapidity data~\cite{CDF:ZRAP_2009}  allows a robust determination of the $d$-valence distribution particularly at high-$x$ where 
previously the functional forms were rather unstable. The new data has allowed the number of constraining 
parameters to be increased and a more robust distribution to be obtained~\cite{MSTW:2009}. The variance on the distribution is now 
larger but believed to be a more reliable estimate of the uncertainty compared to previous fits.

$W^+$ bosons at the Tevatron are preferentially boosted along the incoming proton direction since the $u$-valence 
quark carries on average more momentum than the $d$-valence quark. A measurement of the $W$ charge asymmetry 
as a function of rapidity therefore provides constraints on the ratio of the $d$ and $u$-valence quarks. 
Some discrimination between valence and sea quark 
contributions can be obtained by measuring the charge asymmetry as a function of the lepton $E_T$ since the sea 
quark contribution is enhanced at low $E_T$. Older measurements~\cite
{CDF:WASYM_2005,D0:WASYM_MUON_2008,D0:WASYM_ELEC_2008}
 have used the lepton charge asymmetry but recently CDF~\cite{CDF:WASYM_2009_W} have unfolded the 
measurements back to the $W$ rapidity which in principle provides more information since the PDF information is not 
convoluted with the $V-A$ decay structure. The data is weighted by the two solutions of $Y_W$ based on kinematic 
constraints informed by the MC. The unfolding is an iterative one to remove the dependence on the MC input 
parameters, particularly the PDFs. MSTW have not incorporated the latest data owing to apparent inconsistencies between the latest CDF 
and \dzero\ data and because it was not possible to fit the data well, particularly at high rapidity and high $E_T$. This has 
prompted a flurry of forensic work to try and understand the issues with the poor PDF fits. CDF has taken its $W$ 
charge asymmetry measurement and converted it to a lepton charge asymmetry to facilitate comparisons with 
\dzero. The two datasets are broadly in agreement. The theoretical predictions of the asymmetry have been studied and it 
has been found that MC@NLO~\cite{MC_NLO} provides, somewhat surprisingly, a better description of the data than 
the resummed NLO RESBOS~\cite{RESBOS:LADINSKY_YUAN,RESBOS:1995,RESBOS:1997,RESBOS:2003} 
prediction and that while NNLO effects are small they are not negligible, particularly at the highest rapidities. MSTW~
have also investigated~\cite{MSTW_SHADOWING} the shadowing corrections that are applied to the fixed-target 
structure function data which also constrain the $d/u$ quark ratio. They find that the worst fit is obtained by using the 
default behaviour which is a shadowing correction at low-$x$. A better description of the data is obtained when no 
correction is applied (as CTEQ does) or if the correction is applied at high- as well as at low-$x$. It is clear that there 
are still issues to be resolved before the latest CDF and \dzero\ W asymmetry measurements can successfully be 
incorporated in the PDF fits.

\section{Di-boson Cross Sections and Anomalous Gauge Couplings}
The study of di-bosons at the Tevatron is interesting for two reasons. Firstly beyond the SM (BSM) physics would likely 
manifest itself in anomalous couplings resulting in enhanced production rates and secondly di-boson production are 
generally backgrounds to the Higgs signature. The lepton decay channels present the cleanest mode of identifying 
di-bosons, albeit at the expense of a small branching fraction. CDF has recently remeasured~\cite{CDF:WW} the $WW
$ cross section in the leptonic decay mode and obtained $\sigma_{WW} = 12.1^{+1.8}_{-1.7}$~pb in good agreement 
with the SM prediction~\cite{MCFM_WW_ZZ} of $11.7 \pm 0.7$~pb. It has also obtained limits (for $\Lambda = 1.5
$~TeV) on the anomalous couplings:
$\lambda^Z < 0.16, \Delta{g_{1}^{Z }} < 0.34$ and $\Delta{\kappa^{\gamma }} < 0.72$.  CDF has also observed the 
$ZZ$ signature in the leptonic mode with a significance of greater than $5\sigma$ for the first time, allowing a cross 
section of $\sigma_{ZZ} = 1.56^{+0.8}_{-0.63} (\rm stat.) \pm 0.25 (\rm syst.)$~pb to be determined which again is in good
agreement with the SM prediction of $1.4 \pm 0.1$~pb~\cite{MCFM_WW_ZZ} .

In the recent analyses the emphasis has been to identify $WZ/ZZ/WW$ decays where one of the bosons decays 
hadronically. CDF has used two different 
selection methods for this purpose: the first requires \met\ and two jets and is a cut-based analysis and the second 
requires a charged lepton, \met\ and two jets and is a log-likelihood based analysis. The second selection clearly 
precludes the $ZZ$ mode. In both cases signals of $\sim$ 1,500 events are observed with significances in excess of 5
$\sigma$ with cross sections in good agreement with the SM.  For example the cut based analysis measures a cross 
section of $18.0 \pm 2.8 (\rm stat.) \pm 2.6 (\rm syst.)$~pb compared to the SM prediction of $16.8 \pm 0.5$~pb.
The successful demonstration from the di-boson samples that $W$ and $Z$ bosons can be successfully identified from the 
hadronic decay mode is particularly valuable in the context of Higgs searches since the application of the 
same techniques can be used to enhance the number of possible signal events. 

A search for anomalous couplings has also recently been performed in the $Z\gamma$ channel where such couplings 
would enhance the rate of high $E_T$ photons. A search in the region of $E_T^\gamma > 40$~GeV resulted in 91 
events with a background $\ll$ 1 compared to a SM expectation of 89 events. Limits on anomalous couplings ($h_3, h_4$)
which physically correspond to anomalous electric-dipole and magnetic-quadrupole moments of the $Z$ 
have been determined to be: $|h_3| < 0.037, |h_4| < 0.0017$ which are already significantly better than the LEP limits and which 
are expected to improve by approximately a factor of two when the $Z \rightarrow \nu\overline{\nu}$ channel is 
incorporated into the analysis.

\section{Beyond the Standard Model Constraints}
In addition to the analysis of di-bosons, the large number of single $Z$ and single $W$ events can be used to  
probe for BSM physics. A measurement of the forward backward asymmetry of the decay leptons from 
$Z$ decays ($A_{FB}$) is sensitive to the presence of new $Z$ bosons and the rare decay $W \rightarrow \pi\gamma$ is expected to 
be enhanced in the presence of BSM interactions. CDF has measured $A_{FB}$ up to masses of 500 GeV/$c^2$ using 
4.1~fb$^{-1}$ of integrated luminosity and found excellent agreement with the SM which will subsequently be exploited 
to measure $\sin\theta_W$ to a precision that, with the full Tevatron dataset ($\sim 10$~fb$^{-1}$), should eclipse the 
LEP measurement. In the SM the branching ratio for the rare decay $W \rightarrow \pi\gamma$ is $\sim 10^
{-6}-10^{-8}$ and is expected to be enhanced by BSM interactions. No evidence for a significant enhancement has 
been observed from 4.3~fb$^{-1}$ of integrated luminosity and an upper limit (at 95\% confidence level) of $6.4 \times 10^{-5}$ has been placed on the branching fraction.

\section{Outlook}
The Tevatron is performing extremely well and is presently delivering integrated luminosities in excess of 50~pb$^{-1}$ per week 
and is expected to deliver 10~fb$^{-1}$ before the nominal end of the run in September 2011. This will result in 
electroweak samples far larger than those expected from the first LHC run and significant improvements over the 
analyses presented here are to be expected with increases by factors of at least two in statistical precision across all analyses 
which are for the most part still statistically limited. CDF expects to measure the $W$ mass to a precision of better 20~MeV/$c^2$ which in 
conjunction with a top mass measured to 1~GeV and the central values remaining the same offers the tantalising 
prospect of establishing an upper limit on the Higgs mass (at 95\% confidence) level) below the LEP2 direct search limit of 114~GeV/$c^2$. 
Anomalous couplings and rare decays will be probed with increased precisions and improved PDF constraints are 
expected as the theoretical treatment is refined. The future for electroweak physics at CDF is a rosy one.

\section*{Acknowledgments}
I would like to thank the organizers of the Lake Louise Institute for a supremely well organized, stimulating and 
enjoyable conference in an stunning location and my congratulations to the host nation in winning the Ice Hockey 
Olympic gold medal. I would also like to thank the STFC for their financial support that enabled me to present results 
at this conference.

\section*{References}
\bibliography{lake-louise-iop}

\begin{thebibliography}{10}

\bibitem{CDF:ZRAP_2009}
CDF collaboration, T.~Aaltonen {\em et~al.},
\newblock (2009), arXiv:0908.3914v4.

\bibitem{MSTW:2009}
A.~D. Martin, W.~J. Stirling, R.~S. Thorne, and G.~Watt,
\newblock Eur. Phys. J. {\bf C63}, 189 (2009).

\bibitem{CDF:WASYM_2005}
CDF collaboration, D.~Acosta {\em et~al.},
\newblock Phys. Rev. D {\bf 71}, 051104 (2005).

\bibitem{D0:WASYM_MUON_2008}
D0 collaboration, V.~M. Abazov {\em et~al.},
\newblock Phys. Rev. D {\bf 77}, 011106 (2008).

\bibitem{D0:WASYM_ELEC_2008}
D0 collaboration, V.~M. Abazov {\em et~al.},
\newblock Phys. Rev. Lett. {\bf 101}, 211801 (2008).

\bibitem{CDF:WASYM_2009_W}
CDF collaboration, T.~Aaltonen {\em et~al.},
\newblock Phys. Rev. Lett. {\bf 102}, 181801 (2009).

\bibitem{MC_NLO}
S.~Frixione and B.~R. Webber,
\newblock Journal of High Energy Physics {\bf 2002}, 029 (2002).

\bibitem{RESBOS:2003}
F.~Landry, R.~Brock, P.~M. Nadolsky, and C.-P. Yuan,
\newblock Phys. Rev. D {\bf 67}, 073016 (2003).

\bibitem{RESBOS:1995}
C.~Bal\'azs, J.~Qiu, and C.-P. Yuan,
\newblock Physics Letters B {\bf 355}, 548  (1995).

\bibitem{RESBOS:1997}
C.~Bal\'azs and C.-P. Yuan,
\newblock Phys. Rev. D {\bf 56}, 5558 (1997).

\bibitem{RESBOS:LADINSKY_YUAN}
G.~A. Ladinsky and C.-P. Yuan,
\newblock Phys. Rev. D {\bf 50}, R4239 (1994).

\bibitem{MSTW_SHADOWING}
R.~Thorne,
\newblock {MSTW} {U}pdate,
\newblock {Talk at DIS2010 conference
  (\url{http://www.desy.de/h1zeus/combined_results/benchmark/Thorne.pdf)}},
  2010.

\bibitem{CDF:WW}
CDF collaboration, T.~Aaltonen {\em et~al.},
\newblock Phys. Rev. Lett. {\bf 104}, 201801 (2010).

\bibitem{MCFM_WW_ZZ}
J.~M. Campbell and R.~K. Ellis,
\newblock Phys. Rev. D {\bf 60}, 113006 (1999).

\end{thebibliography}
\bibliographystyle{h-physrev}

\end{document}